\documentclass[twocolumn]{aastex62}

\usepackage{graphicx}
\usepackage{hyperref}
\usepackage{natbib}
\usepackage{color,soul}
\usepackage{amssymb}
\usepackage{amsmath}

\begin{document}

\title{A Spatially Resolved Study of the GRB 020903 Host Galaxy}

\author{Mallory D. Thorp and Emily M. Levesque}
\affil{Astronomy Department, Box 351580, University of Washington, Seattle, WA 98195, USA}
\email{mdt28@uw.edu, emsque@uw.edu}
 
\begin{abstract}

GRB 020903 is a long-duration gamma ray burst (LGRB) with a host galaxy close enough and extended enough for spatially-resolved observations, making it one of less than a dozen GRBs where such host studies are possible. GRB 020903 lies in a galaxy host complex that appears to consist of four interacting components. Here we present the results of spatially-resolved spectroscopic observations of the GRB 020903 host. By taking observations at two different position angles we were able to obtain optical spectra (3600-9000\AA) of multiple regions in the galaxy. We confirm redshifts for three regions of the host galaxy that match that of GRB 020903. We measure metallicity of these regions, and find that the explosion site and the nearby star-forming regions both have comparable sub-solar metallicities. We conclude that, in agreement with past spatially-resolved studies of GRBs, the GRB explosion site is representative of the host galaxy as a whole rather than localized in a metal-poor region of the galaxy.

\end{abstract}


\section{Introduction} \label{sec:intro}
Long-duration gamma-ray bursts (LGRBs) are highly energetic transient events associated with the core collapse of massive progenitor stars \citep{1993ApJ...405..273W}. Given the short lifetimes of their massive progenitor stars, LGRBs occur in star-forming galaxies, with morphologies that are more irregular than normal core-collapse supernovae hosts \citep{2002AJ....123.1111B,2003A&A...400..499L,2004A&A...425..913C,2006Natur.441..463F}. The extreme luminosities of these events allow us to detect them out to extremely high redshifts ($z\sim9.4$,  \citealt{2011ApJ...736....7C}). As a result, LGRBs can serve as excellent probes of the earliest star-forming galaxies in our universe. Consequently, it is important to understand whether the explosion site of an LGRB - including its star formation rate, metallicity, and evolved massive star population - can be treated as representative of both the host galaxy and the local star-forming galaxy population as a whole.

Recent studies have suggested that LGRBs are preferentially located in relatively low-metallicity galaxies \citep{2006ApJ...637..914W,2007ApJ...657..367W,2015A&A...581A.125K}, with LGRB hosts falling below the mass-metallicity and luminosity-metallicity relations for star-forming galaxies out to $z\sim1$ \citep{2008ApJ...683..321F,2009ApJ...702..377K,2010AJ....139..694L,2010AJ....140.1557L,2006ApJ...637..914W}. \citet{2015A&A...581A.125K} identified a population of GRB hosts with greater than solar metallicity (the average GRB host metallicity is log(O/H)+12 $\sim$ 8.5), but this fraction of hosts is small and suggests that some mechanism may be hindering GRB development in high metallicity environments. However, most LGRB studies are limited to global studies of the host galaxies as a whole, studying morphological trends in LGRB localization or obtaining a single spectrum of a small and faint galaxy that represents a galaxy-wide composite of its ISM properties. Host absorption in an LGRB afterglow spectrum is a better representation of the local GRB host environment, but this only samples a single sight-line and is limited to higher redshifts ($z\gtrsim1.5$), where most work is done in the rest-frame UV and is thus difficult to consistently compare to lower-redshift optical studies.

A complete understanding of the environmental dependence of LGRBs relies upon observations that can pinpoint the specific local explosion sites of GRBs and compare them to their larger host galaxies.Unfortunately, GRB rate density peaks at z$\sim$2.5 and decreases by an order of magnitude towards z$\sim$0 \citep{2012ApJ...752...62J,2016ApJ...817....7P}. LGRBs in this z$\leq$1 region preferentially occur in faint, low mass galaxies \citep{2015A&A...581A.102V}. Consequently, LGRB hosts that are close enough or large enough (in projection) for spatially resolved studies are rare, and only five have been studied so far \citep{2016IAUFM..29B.267L}. The closest such galaxy, the host of GRB 980425 at $z=0.0085$, has been studied using integral field unit spectroscopy \citep{2008A&A...490...45C,2017A&A...602A..85K}. \citet{2014MNRAS.441.2034T} and  \citet{2017arXiv170405509I} performed similar IFU analysis using the Very Large Telescope (VLT), on the hosts of GRB 060505 and GRB 100316D respectively; the former being an update to a previous long slit study by \citet{2008ApJ...676.1151T}. All provided an extensive series of spatially resolved spectra and complex metallicity maps, demonstrating that GRBs form in regions that appear to be slightly more metal-poor than the global metallicity but agree to within the uncertainties of the metallicity diagnostics \citep{2008ApJ...681.1183K}.

The remaining hosts have been studied using long-slit spectroscopy, positioning the slit to capture multiple distinct regions within the presumed host complex (the hosts of GRB 120422A \citealt{2012ApJ...758...92L, 2014A&A...566A.102S}, and GRB 020819B \citealt{2010AJ....140.1557L}). However, a more recent study of the GRB 020819B host using the Multi Unit Spectroscopic Explorer (MUSE) on the VLT revealed the host galaxy to be a foreground galaxy unassociated with the GRB \citep{2017MNRAS.465L..89P}. This further demonstrates the advancements provided by IFU studies over traditional long-slit spectroscopy. The host galaxy of the ultra-long GRB 130925A at $z=0.347$ has also been studied with spatially-resolved longslit spectroscopy \citep{2015A&A...579A.126S}, although it is unclear whether the ultra-long class of GRBs should be treated as phenomenologically distinct from the general LGRB population (e.g. \citealt{2014ApJ...781...13L}). In all studies where the host is successfully resolved into multiple components, the observations conclude that the explosion site metallicity is representative of the galaxy as a whole \citep{2012ApJ...758...92L}.

GRB 020903 (sometimes also referred to as XRF 020903 given the soft but X-ray rich spectrum; \citealt{2001grba.conf...16H}) was originally detected on 2002 September 3.421 UT on the Wide-Field X-Ray monitor and the Soft X-Ray Camera on the High Energy Transient Explorer-2 (HETE-2) \citep{2004ApJ...606..994S}. Originally the afterglow went undetected \citep{2002GCN..1531....1T,2002GCN..1533....1P,2002GCN..1535....1P,2002GCN..1537....1U}, given the interference from a nearby galaxy, and by the time \citet{2004ApJ...606..994S} detected an optical and radio afterglow (using the Palomar Observatory and the Very Large Array) the host galaxy dominated the optical spectrum. The radio afterglow was successfully observed and 1000 times more luminous than that of a Ibc supernova, indicating the presence of a GRB afterglow \citep{2004ApJ...606..994S}. An optical rebrightening was observed approximately 25 days after the initial detection, indicating that a supernova might be associated with the GRB \citep{2002GCN..1554....1S,2006ApJ...643..284B}. Given the similarities between the light curve and spectral distribution curve of this afterglow to those of SN 1998bw, a supernova is the only plausible source for this GRB \citep{2006ApJ...643..284B}. The first study of the LGRB host environment showed it to be a low-metallicity starburst galaxy and concluded that it appeared to have at least 4 interacting components \citep{2004ApJ...606..994S, 2002GCN..1761....1L}. The explosion site itself has been previously observed spectroscopically and found to have one of the lowest metallicities measured for a GRB host, with log(O/H)+12 $\sim$ 8.0 \citep{2010AJ....139..694L}, as well as features of a significant Wolf-Rayet star population \citep{2006A&A...454..103H, 2010A&A...514A..24H}. However, the other three bright components of the host complex have never been spectroscopically observed, and it is unclear whether these are indeed the results of an interaction or merger (as proposed by \citealt{2004ApJ...606..994S, 2005ApJ...633...29C, 2007ApJ...657..367W}) or whether the larger host complex contains an active galactic nucleus as proposed by \citet{2002PASP..114..587G}.

Here we present a spatially-resolved study of the GRB 020903 host galaxy, using long-slit spectra of multiple locations within the host complex. Observations and reductions are discussed in Section 2. We determine redshifts and ISM properties for these regions (Section 3) and discuss our results and comparisons with previous work in Section 4.

\section{Observations \& Reductions} \label{sec:methods}
\subsection{Observations} 
We were allocated 4.5 hours of observing time through the Gemini Fast Turnaround program on the Gemini Multi-Object Spectrograph on Gemini South. Observations were carried out in queue mode during November and December of 2016 (see Table 1), with an image quality of $\ge$70$\%$\footnote[1]{Our image quality criteria correspond to a point source FWHM of $\le0.75''$ in the $r$-band; see \texttt{\url{http://www.gemini.edu/sciops/telescopes-and-sites/observing-condition-constraints\#ImageQuality}} for a complete discussion of Gemini image quality criteria.} and a mean airmass of 1.19.

\begin{figure*}
    \centering
    \includegraphics[width=0.8\textwidth]{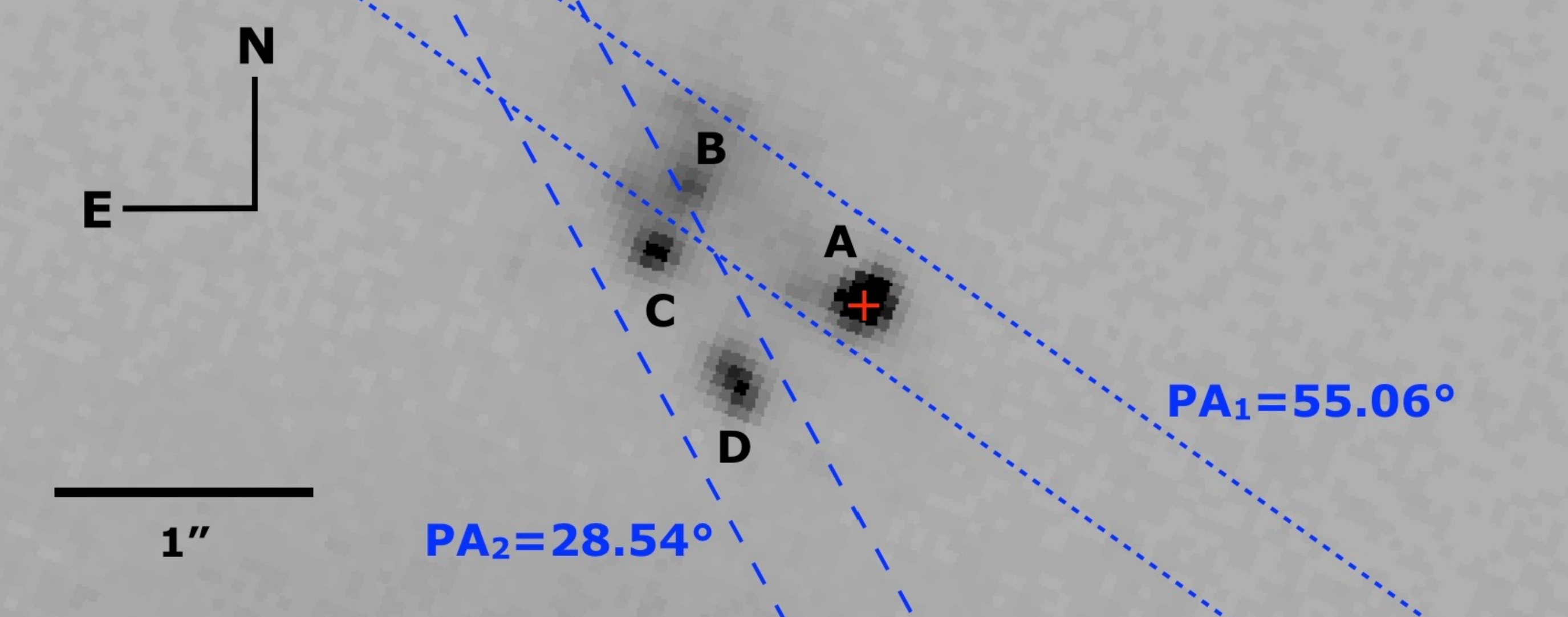}%
    \caption{The host complex of GRB 020903, imaged using the F606W filter on ACS with HST. The four key regions identified in the host complex (ABCD) are labeled; the explosion site associated with region A is marked with a red cross. The two slit position angles used in our observations are illustrated with dotted (PA$_1$=55.06$^{\circ}$, spanning the A and B regions) and dashed (PA$_2$=28.54$^{\circ}$, spanning the C and D regions) lines indicating the width of the 0.5$\arcsec$ slit.}%
    \label{fig:HST}%
\end{figure*}

Our spectra of the GRB 020903 host complex were taken with the 0.5$''\times$330$''$ slit, using the B600 grating centered at 5000\AA\ and the R400 grating (with the GG0405 blocking filter) centered at 7000\AA\ for a total wavelength coverage of $\sim$3600\AA\ to 9000\AA\ (with the grating centers shifted by +50\AA\ for half of our observations to avoid the detector gaps). Given the irregular nature of the GRB 020903 host complex, we observed spectra of the galaxy at two different specific position angles: 55.06$^{\circ}$ to observe the A (GRB explosion site) and B regions of the host, and 28.54$^{\circ}$ to observe the C and D regions (see Figure \ref{fig:HST}). As a result of this technique, some of our observations were taken well off of the parallactic angle (up to $\sim$62$^{\circ}$). Since this technique was the only means of ensuring that we could observe two regions of the host complex in a single observation, it rendered flux calibration of our data impractical.

To ensure that we did indeed capture the GRB host galaxy in the slit, we first centered the slit on nearby bright stars that would would place both the centered star and the specified regions of the host complex on the slit at the appropriate position angle. We then confirmed our slit placement in the raw 2D images, successfully identifying continua from the host spectra and other intervening objects at the appropriate positions along the slit.

A full summary of observing dates and configurations is given in Table 1; each configuration was observed for a total exposure time of 1800s. We also obtained standard quartz lamp and CuAr arc lamp observations for flatfield and wavelength calibrations. The pixel scale of the GMOS-S detector is $0.08''$/pix and the data were reading out in 2x2 binning mode, giving our data an image scale of $0.16''$/pixel.

\begin{deluxetable}{lcccc}
\tabletypesize{\scriptsize}
\tablecaption{Summary of observations}
\tablewidth{0pt}
\tablehead{
\colhead{Date (UT)} & \colhead{PA ($^{\circ}$)} & \colhead{Grating} & \colhead{$\lambda_{c}$ (\AA)}
}
\startdata
2016 Nov 5 & 55.06 & B600 & 5000 \\ 
  & 55.06 & B600 & 5050 \\ 
2016 Nov 28 & 55.06 & R400 & 7000  \\ 
  & 55.06 & R400 & 7050 \\
2016 Dec 2 & 28.54 & R400 & 7000 \\ 
  & 28.54 & R400 & 7050 \\ 
2016 Dec 20 & 28.54& B600 & 5000 \\
\enddata
\label{table:Obs}
\end{deluxetable}

\subsection{Reduction}
The data were reduced using IRAF\footnote[2]{IRAF is distributed by the National Optical Astronomy Observatories, which are operated by the Association of Universities for Research in Astronomy, Inc., under cooperative agreement with the National Science Foundation.}, primarily using the \texttt{GMOS} package tailored specifically for reduction of Gemini spectrograph data and including standard routines for bias correction, flatfielding, and cosmic ray removal. Initially, a single spectrum was extracted from each file using the IRAF task \texttt{apall} and a 10-pixel ($1.6''$) aperture, large enough to yield composite spectra of regions A+B (for spectra taken at a position angle of 55.06$^{\circ}$) and C+D (for spectra taken at a position angle of 28.54$^{\circ}$).

\begin{figure*}[h!]
    \centering
        \includegraphics[width=.4\textwidth,angle=270]{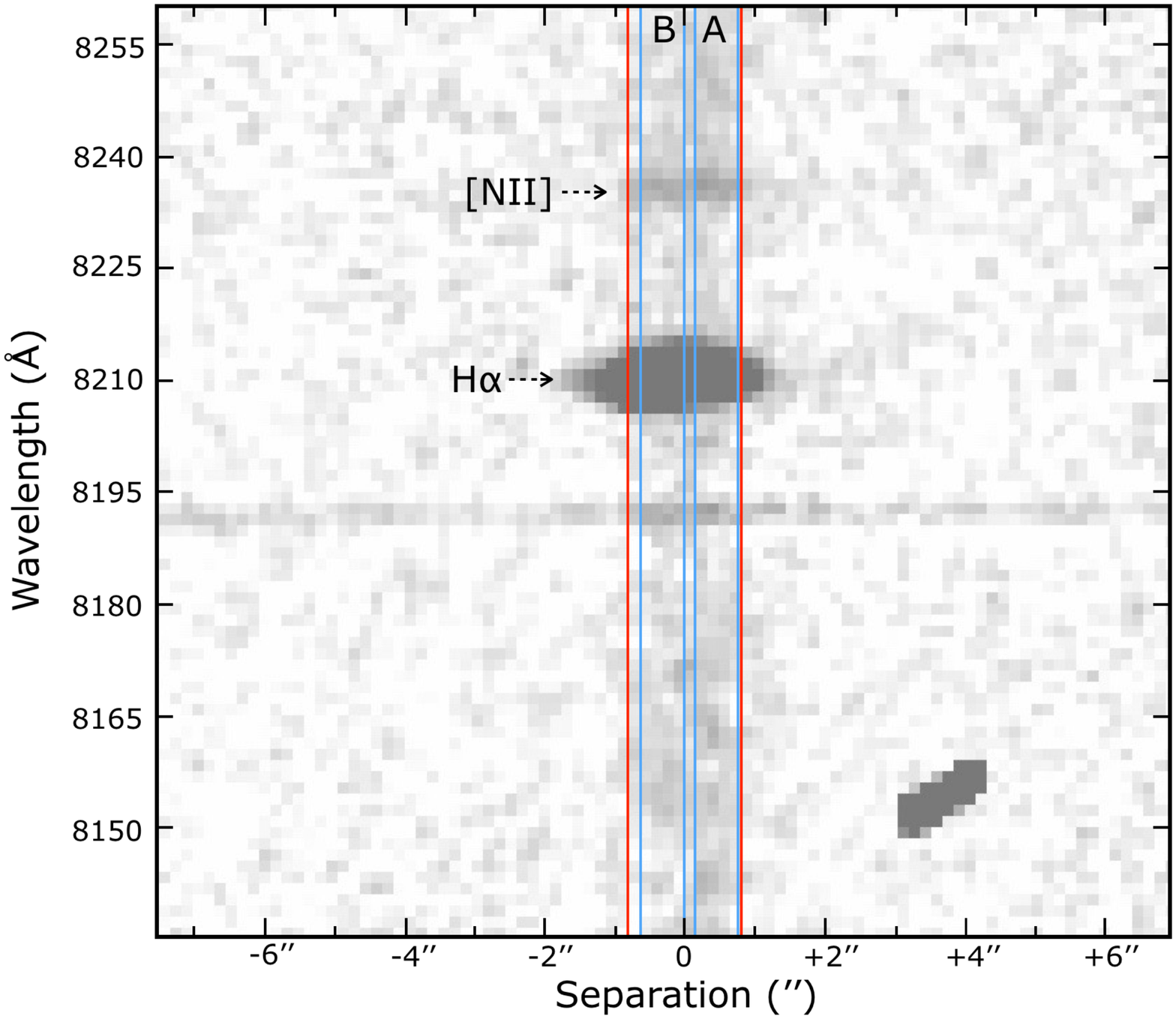} \\
        \includegraphics[width=.4\textwidth]{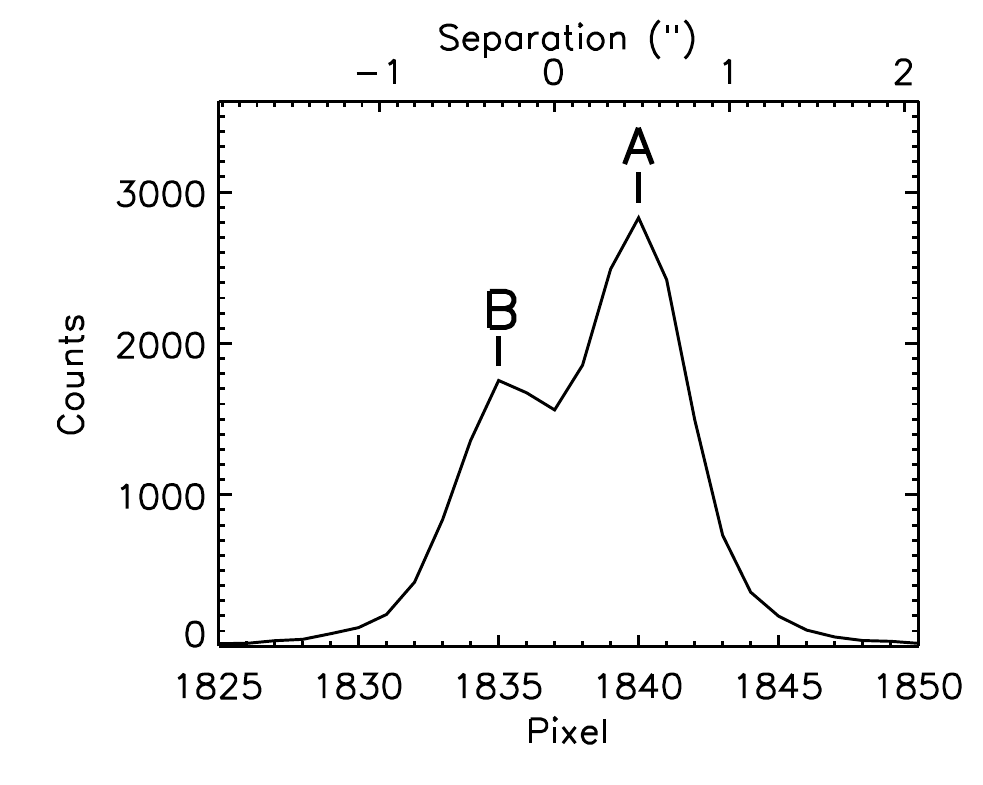} \\
        \includegraphics[width=.4\textwidth]{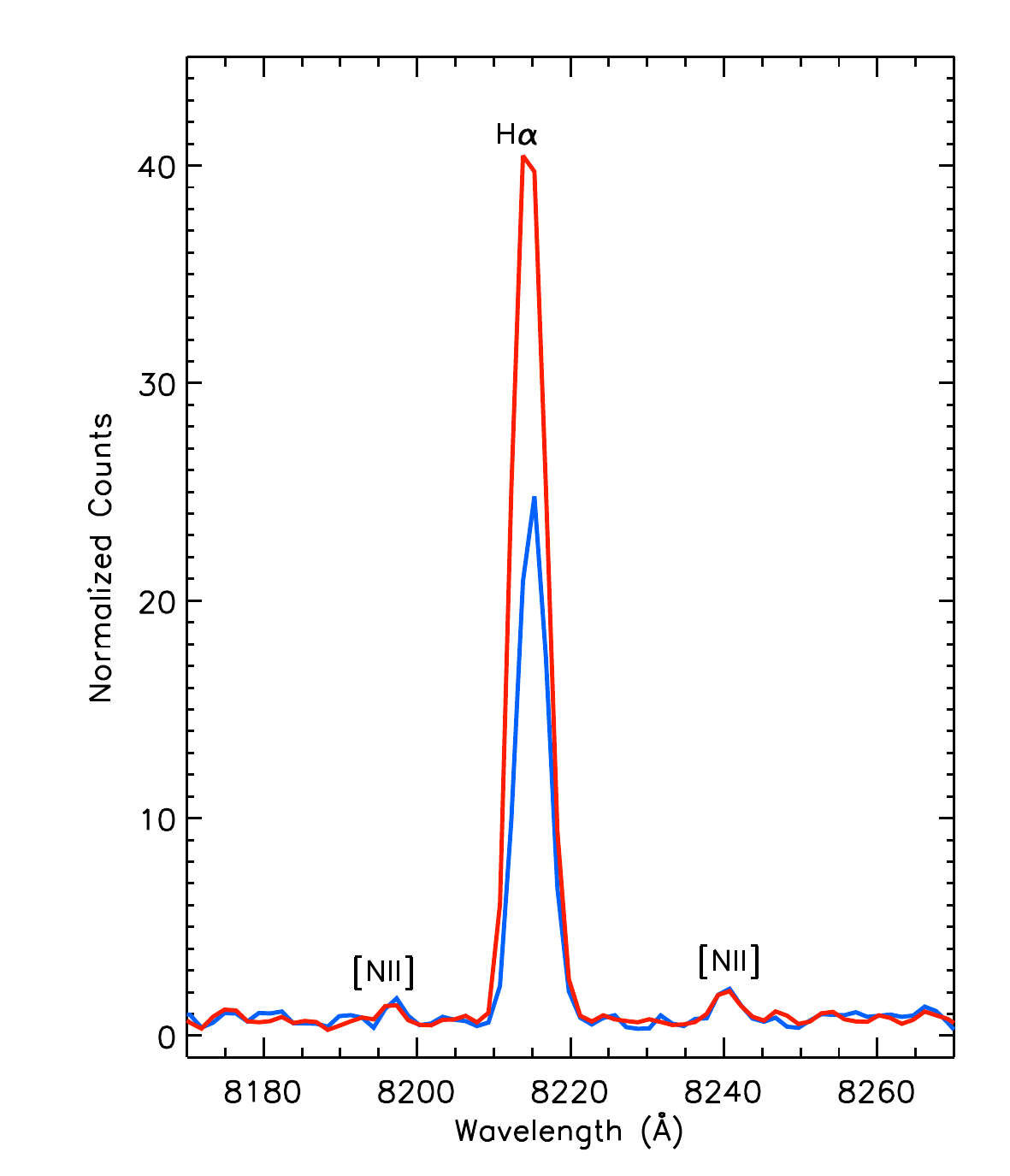}
        \caption{Extraction of individual spectra for the A (explosion site) and B regions of the GRB 020903 host complex. {\bf Top}: raw 2D ds9 image (vertical dispersion direction) of an observation taken at a position angle of 55.06$^{\circ}$; the continuum from the A and B regions of the host complex is visible, as are the bright H$\alpha$ and [NII]$\lambda$6584 emission features. Extraction regions for the composite A+B spectrum (red) and the individual A and B spectra (blue) are marked. {\bf Center}: H$\alpha$ line profile extracted from the above image; separate peaks corresponding to the two regions are labeled. {\bf Bottom}: Comparison of the individual continuum-normalized spectra extracted for region A (red) and B (blue); positions of the H$\alpha$ and [NII]$\lambda$6548 lines at $z=0.251$ are marked. Note the changes in the strengths of the H$\alpha$ and [NII]$\lambda$6584 features.}
    \label{fig:Red_process}
\end{figure*}

In addition, we extracted an H$\alpha$ line profile in the spatial direction (also using the \texttt{apall} optimal extraction algorithm) for our A+B observations. The line profile revealed two spatially-distinct peaks in our A+B observations, f$0.8''$ (5 pixels) apart, corresponding to separate spectra from the A and B regions (Figure \ref{fig:Red_process}) and discernible given the maximum FWHM image quality criteria required for executing our observations (FWHM$\le0.75''$ in $r$); no similar distinction could be discerned in the line profiles for the C+D observations. To extract individual spectra of regions A and B within the host, we performed a second extraction centered on each peak with two smaller apertures of 4 pixels ($0.64''$) to minimize blending from the two regions in the individual spectra. However, given the close proximity of the apertures (the peak of Region A is separated from the Region B aperture by only $0.64''$ and vice versa) some contribution to the spectrum of each region from its neighbor is still to be expected. Assuming Gaussian brightness profiles for both regions we estimate a contribution of $\lesssim$10\% from Region B in the Region A spectrum and $\lesssim$20\% from Region A in the Region B spectrum.

Beginning on 2016 Sept 30 a new bias structure appeared on the GMOS-S detector following a full thermal cycle of the dewar. This most notably included broad bright vertical fringes and 1-pixel horizontal stripes in CCD2 and CCD3 that could not be reliably subtracted in full, and a significant increase in the detector noise (as much as five times higher than unaffected areas in CCD2). A solution for removing this bias structure was not found during engineering tests, but the problem spontaneously resolved following an unscheduled thermal cycle of the instrument on 2017 Feb 21. All of our observations were taken while this bias structure and excess noise were present on the detector, including biases (Gemini support advised using contemporaneous bias frames to minimize the effects of this problem). As the central chip, CCD2, was the most heavily affected, this made all data between $\sim$4800-5500\AA\ (for our blue spectra) and $\sim$6000-7800\AA\ (for our red spectra) unsuitable for line measurements. Our grating configurations did provide clean wavelength coverage below 4800\AA\ and between $\sim$5500-6500\AA\ (for the B600 grating centered at 5050\AA) and between $\sim$7800-9000\AA\ (for the R400 grating centered at 7050\AA), successfully capturing the critical [OII]$\lambda$3727, H$\beta$, [OIII]$\lambda$4959, [OIII]$\lambda$5007, H$\alpha$, and [NII]$\lambda$6584 features at the presumed GRB 020903 host redshift of $z=0.251$ in the clean regions of CCD3 (see Section 3.1 for further discussion).

To avoid compounding the effects of the CCD2 and CCD3 bias structures, we reduced each individual observation of the GRB 020903 host complex separately. The effects of the bias structure precluded effectively combining spectra taken with two different gratings or two different central wavelengths, since shifting bias structure and chip gap effects would inconsistently impact our emission features across different configurations. As a result, our final analyses were conducted using only the B600 $\lambda_c$=5050\AA\ (where we detect the [OII], H$\beta$, and [OIII] emission features of the GRB host) and R400 $\lambda_c$=7050\AA\ spectra (where we detect the H$\alpha$ and [NII] emission features of the GRB host) for the A and B regions. Due to a slight misalignment of the position angle, B600 spectra were not obtained for the C+D region, only R400 spectra (where we detect the H$\beta$, [OIII]$\lambda$4959, H$\alpha$ and [NII] emission features). [OIII]$\lambda$5007 is lost in the chip gap, and the C+D region was dimmer than the A+B region, making [OII]$\lambda$3727 indistinguishable from the continuum.

\begin{figure}
    \centering
        \includegraphics[width=\columnwidth]{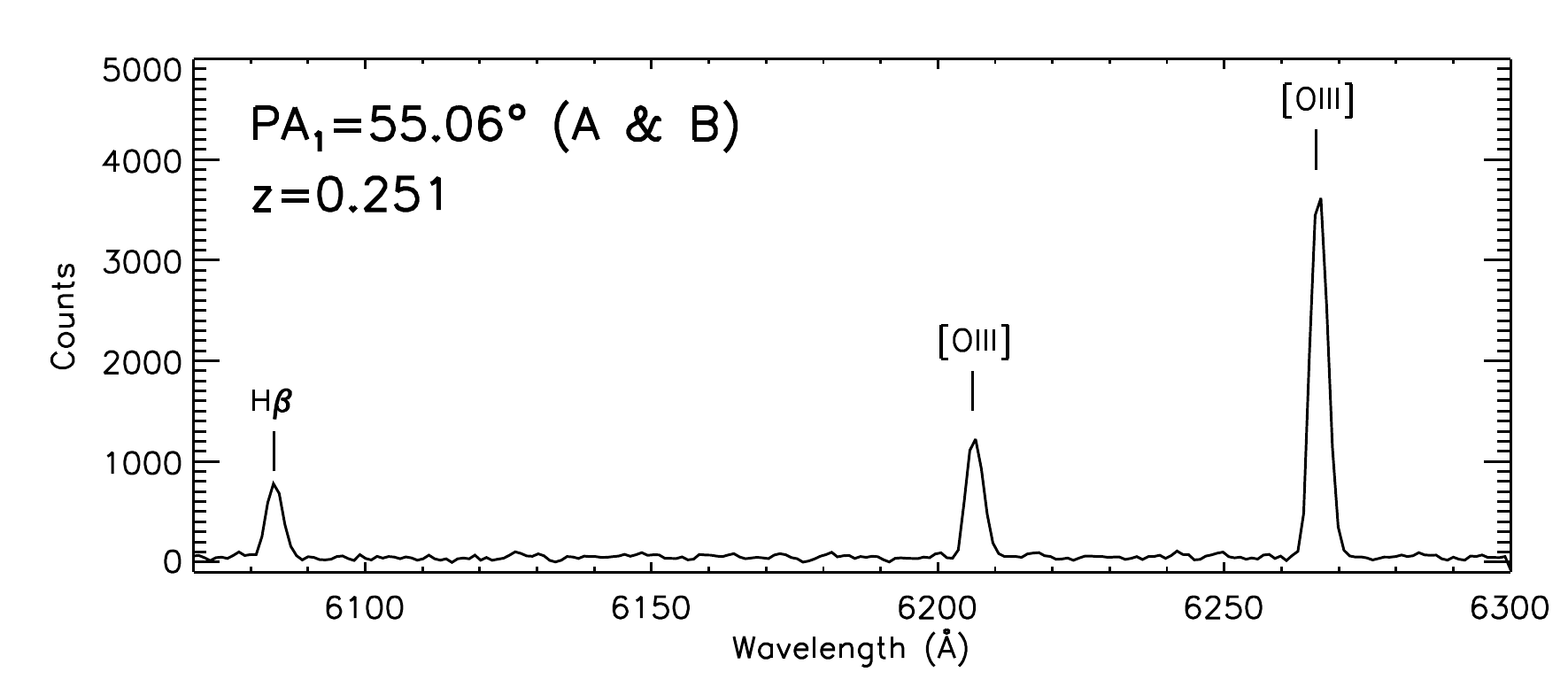}
        \includegraphics[width=\columnwidth]{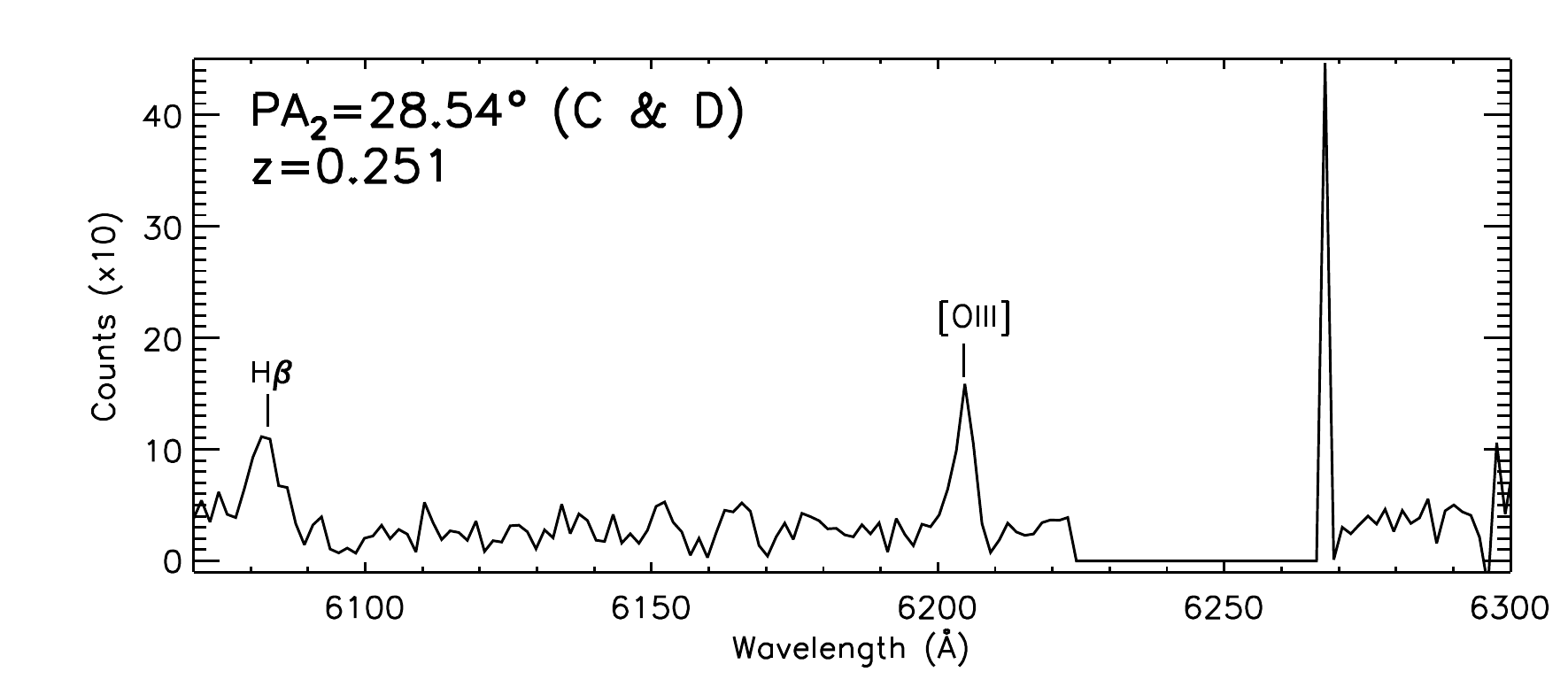}
    \caption{Composite spectra of the A+B regions (top) and C+D regions (bottom) for the GRB 020903 host galaxy. Note the chip gap present in the C+D spectrum at $\sim$6225\AA-6265\AA\ due to insufficient wavelength coverage at PA$_2$, which prevents measuring a clean emission line profile and equivalent width for the [OIII]$\lambda$5007 feature.}
    \label{fig:zoom}
\end{figure}

Figure \ref{fig:zoom} shows sections from the composite spectra of regions A+B and C+D, confirming that the redshift of these regions is z=0.251 and demonstrating how the C and D regions are significantly dimmer. This comes into significant effect when we consider any contributions from the C+D spectrum in the A+B spectrum and vice versa. Regions B and C in particular are quite close together; one could thus expect potential contributions from each region in the other's spectrum. It is possible that region C+D is not strongly star-forming and thus does not contribute significant emission signatures to the A and B spectra. It is also possible that contribution of region B is too weak to be discerned in the C+D spectrum; however, this explanation seems unlikely given the relative brightness of region B in Figure \ref{fig:HST}. It is possible that the image quality during these observations was sufficient to avoid significant blending. Unfortunately, Gemini notes only the image quality {\it percentile} for individual exposures (corresponding here to FWHM$\le0.75''$ in $r$), so the precise FWHM during these observations cannot be confirmed. More likely there is some blending between regions, especially between regions C and B, that is unavoidable.

\section{Analysis} \label{sec:Analysis}

 Emission line equivalent widths were measured by continuum normalizing the spectra using the task \texttt{continuum} and using the \texttt{splot} task in the \texttt{kpnoslit} package to integrate the pixel values of each spectral feature from the linear continuum. Table \ref{table:Eqw} summarizes the average line widths measured in regions A, B, and C+D, corrected for reddening and redshift effects.

\begin{deluxetable}{lccc}
\tabletypesize{\scriptsize}
\tablecaption{Average dereddened equivalent widths of emission lines \label{table:Eqw}}
\tablewidth{0pc}
\tablecolumns{4}
\tablehead{\colhead{Wavelength} & \colhead{Region A (\AA)} & \colhead{Region B (\AA)} & \colhead{Region C+D (\AA)}}
\startdata
$[OII] \lambda$3727  & 41.57$\pm$1.96  & 17.20$\pm$0.86 & NA\\
H$\beta$             & 59.19$\pm$1.80  & 30.14$\pm$1.12 & 7.43$\pm$1.00\\
$[OIII] \lambda$4959 & 99.76$\pm$3.44  & 68.16$\pm$2.11 & 10.79$\pm$2.05\\
$[OIII] \lambda$5007 & 321.7$\pm$10.71 & 218.8$\pm$7.35 & NA\\
H$\alpha$            & 305.2$\pm$8.02  & 167.6$\pm$4.38 & 9.60$\pm$0.91\\
$[NII] \lambda$6584  & 9.42$\pm$0.63  & 8.64$\pm$0.51 & 0.51$\pm$0.10\\
\enddata
\end{deluxetable}

The spectra were dereddened using the Pyastronomy package \texttt{pyasl.unred}.\footnote[3]{\texttt{\url{https://github.com/sczesla/PyAstronomy}}} $E(B-V)$ values were calculated using the \citet{1989ApJ...345..245C} reddening law and the H$\alpha$/H$\beta$ line ratio, assuming case B recombination (H$\alpha$/H$\beta$ = 2.87) with an effective temperature of $10^{4}$ K and $n_{e}\sim10^{2}-10^{4}$ $cm^{-3}$ following \citet{osterbrock1989astrophysics}. We measured an $E(B-V)$ = 0.03$\pm$0.03 in the A explosion site region, in good agreement with \citet{2010AJ....139..694L}. Region B had a slightly higher $E(B-V)$ = 0.10$\pm$0.03, as did Region C+D with $E(B-V)$ = 0.16$\pm$0.02.

After correcting the spectra for reddening effects, we proceeded to compare the metallicities of the three regions, using B and C+D as a comparative host regions to the GRB site A. We calculated metallicities using the R$_{23}$ diagnostic presented in \citet{2004ApJ...617..240K} for regions A and B, which was calibrated using emission line equivalent widths and is thus well-suited for use with our data. Unfortunately the $[OII] \lambda$3727 is not available for the C+D region. To compensate, we also include metallicities calculated using the O3N2 and N2 diagnostic calibrations of \citet{2004MNRAS.348L..59P}, though we note that these diagnostics are typically applied to data sets where line fluxes are available. The $[OIII] \lambda$5007 is not available for the C+D spectrum, so we adopt a $[OIII]\lambda$5007 / $[OIII]\lambda$4959 ratio of 3 and use this to estimate a $[OIII]\lambda$5007 line.

The resulting metallicities are given in Table \ref{table:Metallicity}. While there are offsets on the order of $\sim$0.1-0.2 dex between the different diagnostics (in agreement with the diagnostic offsets found by \citealt{2008ApJ...681.1183K}), we measure a $O3N2$ metallicity of log(O/H)+12=8.0$\pm$0.1 for region A and B, and log(O/H)+12=8.1$\pm$0.2 for region C+D. While both the R$_{23}$ and N2 diagnostic suggest that regions B,C, and D may have a slightly higher metallicity, all three diagnostics show that the two regions have comparable metallicities to within the errors (with the systematic errors of the metallicity diagnostic calibrations dominating the sources of error; see \citealt{2008ApJ...681.1183K}).

\begin{deluxetable}{lccc}
\tabletypesize{\scriptsize}
\tablecaption{Metallicities in the GRB 020903 host galaxy \label{table:Metallicity}}
\tablewidth{0pc}
\tablecolumns{4}
\tablehead{\colhead{Region} &\multicolumn{3}{c}{log(O/H) + 12} \\ \cline{2-4}
\colhead{} & \colhead{$R_{23}$} & \colhead{PP04 $N2$} & \colhead{PP04 $O3N2$}}
\startdata
A &8.1$\pm$0.1 &8.1$\pm$0.1 & 8.0$\pm$0.1\\
B &8.3$\pm$0.1 &8.2$\pm$0.1 & 8.0$\pm$0.1\\
C+D &NA &8.2$\pm$0.2 & 8.1$\pm$0.2\\
\enddata
\end{deluxetable}

The rest-frame equivalent width of the H$\beta$ emission line ($W_{H\beta}$) can be used as a diagnostic of the age of the young stellar populations in a star-forming galaxy (e.g. \citealt{1981Ap&SS..80..267D, 2013ApJ...779..170L}). Despite sensitivities to the initial mass function, metallicity, and stellar mass loss rate, $W_{H\beta}$ is predominantly dependent on the evolution of HII regions. $W_{H\beta}$ decreases monotonically with population age, and effects from electron temperature and density are negligible. Therefore, if a zero-age instantaneous burst star formation history is assumed  $W_{H\beta}$ can be used to approximate the {\it typical} age of the galaxy's young stellar population \citep{1986A&A...156..111C,2013ApJ...779..170L}. To clarify, stars of this age dominate the galaxy's continuum, so many young stars within the galaxy have ages equal to or less than the typical young stellar population age. It is also crucial to note the drawbacks of this method, particularly that it adopts a zero-age instantaneous burst star formation history when modeling how $W_{H\beta}$ decreases with age \citep{1986A&A...156..111C}; in reality the star formation history of the galaxy is likely much more complex. Despite these limitations, the age - $W_{H\beta}$ relation still provides a valuable, if approximate, insight on the young stellar population of the host galaxy.

We applied the equations for stellar population age derived by \citet{2010AJ....139..694L}, based on data from \citet{1998ApJ...497..618S}, adopting a model metallicity of Z$\sim$0.004 for all three regions following our own metallicity determination. Region A had a typical young stellar population age of  $4.9\pm{0.1}$ Myr, while region B yielded a slightly older typical young stellar population age of $5.8\pm{0.2}$ Myr. Region C+D had a significantly older typical young stellar population age of $8.9^{+0.7}_{-0.6}$ Myr. This is in agreement with the predicted young ages of GRB progenitors and their parent stellar populations \citep{2002AJ....123.1111B,2007ApJ...660..504B}, as well as predicted masses for GRB progenitors. Adopting the \citet{2013A&A...558A.103G} stellar evolutionary tracks (adopting a sub-solar - log(O/H)+12 $\sim$ 8 - metallicity similar to that of the GRB 020903 host and assuming non-rotating single star evolution), these ages correspond to a zero-age main sequence (ZAMS) mass of $\sim$40M$_{\odot}$ for Region A, as compared to $\sim$32M$_{\odot}$ for Region B and $\sim$20M$_{\odot}$ for the C+D region. This is in good agreement with recent work on the presumed progenitor masses of GRBs: \citet{2007MNRAS.376.1285L} estimated that GRB progenitors must have ZAMS masses significantly higher than 20M$_{\odot}$ (with a model considering only main sequence lifetimes, making this a very conservative lower limit), while \citet{2008ApJ...689..358R} use host galaxy modeling to conclude that progenitors of GRBs have ZAMS masses above $\sim$40$_{\odot}$. The age and inferred ZAMS mass of Region A is thus in excellent agreement with basic predictions for GRB progenitors (though it is worth noting that the effects of rotation and binary interactions will also impact the expected ZAMS mass range of core-collapse progenitors as determined from stellar population ages).

\citet{2010AJ....139..694L} determined a minimum age of 5.8$\pm$0.2 Myr for region A, the explosion site, based on a $W_{H\beta}$ = 31.3. However, it is important to note that the GRB 020903 spectrum presented in \citet{2010AJ....139..694L} was, unusually, taken only about a year after GRB 020903 was detected, on 2003 Oct 7. At this time the light from the explosion site may still be contaminated by the fading core-collapse supernova associated with the GRB (see, for example, \citealt{2006ApJ...643..284B}), which would increase the continuum flux in the blue and lead to an underestimate of the equivalent width for the host's nebular H$\beta$ emission. It should also be noted that blending with other components of the host (or background) galaxies could have complicated an accurate determination of the H$\beta$ equivalent width in Region A given that the \citet{2010AJ....139..694L} spectrum was taken with a $1''$ slit rather than the narrow $0.5''$ slit used here at a position angle of 72.6$^{\circ}$. This set-up could potentially introduce blending with other components of the host complex depending on the seeing conditions.

\section{Discussion and Future Work} \label{sec:Disc}

Overall, our results show that the A region, the explosion site of GRB 020903, is less dusty and may contain a younger massive star population than B, C, and D regions. The young stars and low dust content suggest that Region A is likely the most recent and active site of star formation in the GRB 020903 host complex. This is in agreement with other spatially-resolved host studies of GRB host galaxies that have localized GRBs in the most strongly star-forming regions of their hosts. We also find that the all regions share comparably low metallicities based on the R$_{23}$, O3N2, and N2 diagnostics. This is the most metal-poor host galaxy yet studied with spatially-resolved spectra, and our results agree with similar conclusions for the hosts of GRB 980425, 060505, 100316D, and 120422A, which all support relatively homogeneous metallicity within the galaxy \citep{2008A&A...490...45C,2008ApJ...676.1151T, 2011ApJ...739...23L,2012ApJ...758...92L,2014A&A...566A.102S, 2017arXiv170405509I, 2017A&A...602A..85K}.

\begin{figure}
    \centering
    \includegraphics[width=0.99\columnwidth]{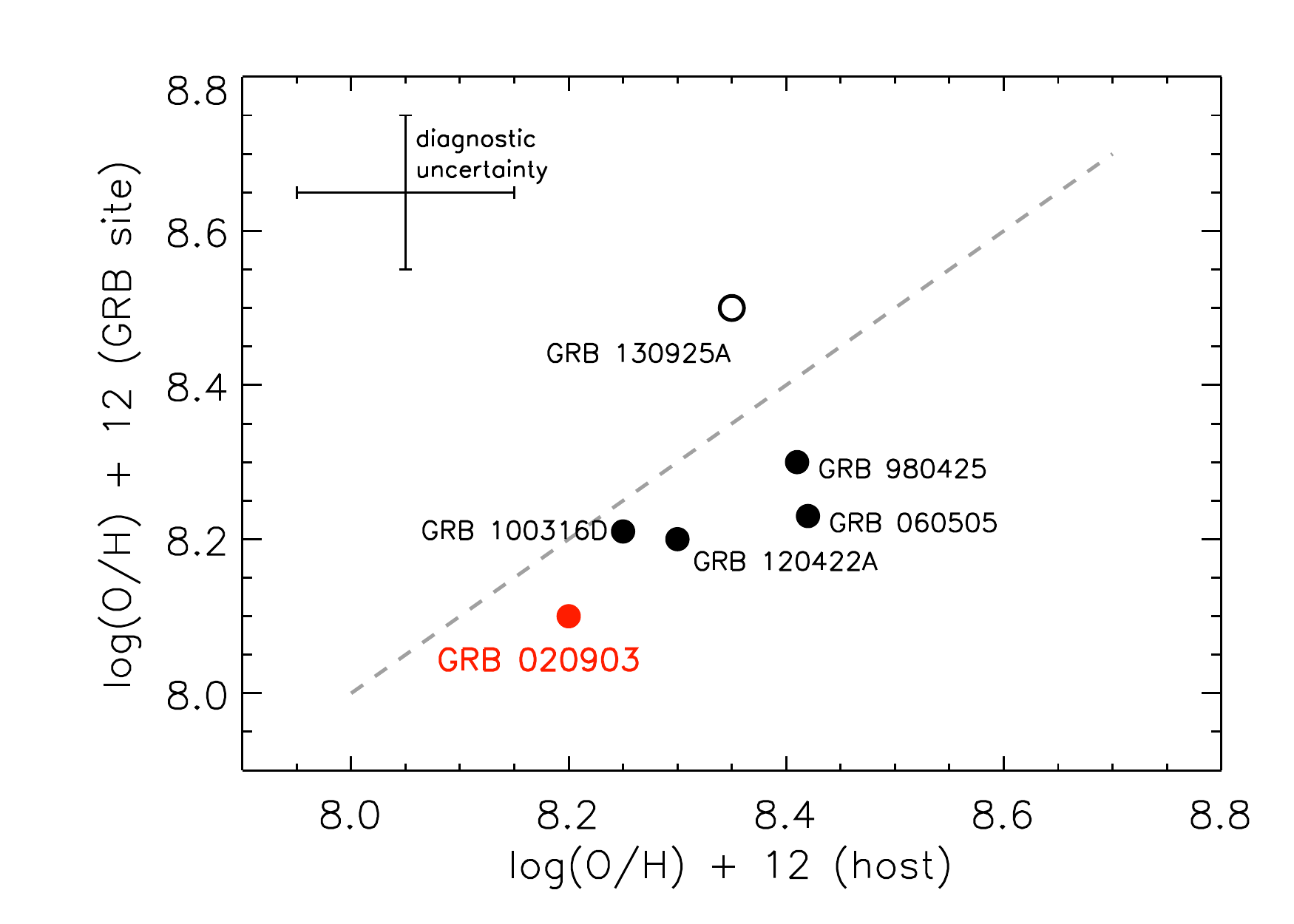}%
    \caption{Updated from Levesque et al.\ (2011);  a comparison of ``host" and ``explosion site" metallicities for LGRBs (filled circles) and the ultra-long GRB 130925A (open circle) as determined from the \citet{2004MNRAS.348L..59P} diagnostics. Our N2 diagnostic results for GRB 020903 are shown in red and compared to metallicities for GRB 980425 \citep{2008A&A...490...45C},  060505 \citep{2014MNRAS.441.2034T}, 100316D \citep{2017arXiv170405509I}, 120422A \citep{2012ApJ...758...92L}, and 130925A \citep{2015A&A...579A.126S}.
    Where site and host metallicities are identical is plotted as a gray dashed line. \label{fig:Fit}}
\end{figure}

 Figure \ref{fig:Fit} illustrates that all six existing studies of spatially-resolved GRB hosts (including the ultra-long GRB 130925A and our work on GRB 020903, but now excluding GRB 020819B based on \citealt{2017MNRAS.465L..89P}) measure explosion site and galaxy metallicities that agree to within the uncertainty of the diagnostics. This further supports evidence that explosion site metallicities can be considered representative of the entire host galaxy, and that GRB host metallicities determined from global spectra can be adopted as acceptable proxies for the natal metallicity of the GRB progenitor.  It is worth noting that at the distances of most GRBs ``site" metallicities remain unresolved on the scale of a kpc or more; for comparison, \citet{2015MNRAS.449.2706N} found that a 500 pc resolution is needed to discern metallicity variations and 100 pc resolution is needed to avoid systemic errors. However, at the typical distance of GRBs kpc resolution is the best attainable resolution for studying potential variations - even at relatively large scales - across their hosts.
 
 This is, of course, based on the still-small sample of spatially-resolved GRB hosts with well-studied ISM properties. There are  other GRB host galaxies where spatially resolved studies of the ISM are possible but have not yet been performed, including GRBs 990705, 011121, 030329, 060218, and 130427A \citep{2016IAUFM..29B.267L}. With a larger sample of spatially-resolved GRB host studies we could draw more concrete conclusions about the precise environments and parent populations that produce GRB progenitors, and how representative these regions are of the galaxies as a whole. These are also compelling future targets for integral field unit (IFU) spectroscopy, which offers the possibility of constructing finer-grained metallicity maps and clearly distinguishing between interacting components and foreground or background regions. \citet{2008A&A...490...45C} and \citet{2017A&A...602A..85K} successfully obtained IFU observations of the very nearby ($\sim$44Mpc) GRB 980425 host galaxy using VIMOS and MUSE (respectively) on the Very Large Telescope; in the future, IFU instruments on the Extremely Large Telescopes should make it possible to extend this work to greater distances and smaller angular sizes, allowing for a larger and more detailed census of GRB explosion sites. 
 
 Finally, in the case of GRB 020903, studies of additional ISM properties such as ionization parameter and star formation history, combined with dynamical studies that can highlight potential past interactions between the A region and the other host regions, will allow us to further characterize the key environmental parameters that led to the birth of the GRB's progenitor star.

\acknowledgments
We wish to thank Trevor Dorn-Wallenstein, David Kann, Locke Patton, Patricia Schady, and Paula Szkody for their valuable advice and feedback on this work. We also wish to thank the anonymous referee for invaluable feedback that has greatly improved the quality of this manuscript. This paper was based on observations (Program ID GS-2016B-FT-7) obtained at the Gemini South Observatory (processed using the Gemini IRAF package) as part of the Gemini Fast Turnaround program. We thank the entire support staff of Gemini Observatory for their efforts and assistance, with particular thanks to Kristin Chiboucas, Mischa Schirmer, Rene Rutten, and Morten Anderson. Gemini is operated by the Association of Universities for Research in Astronomy, Inc., under a cooperative agreement with the NSF on behalf of the Gemini partnership: the National Science Foundation (United States), the National Research Council (Canada), CONICYT (Chile), Ministerio de Ciencia, Tecnolog\'{i}a e Innovaci\'{o}n Productiva (Argentina), and Minist\'{e}rio da Ci\^{e}ncia, Tecnologia e Inova\c{c}\~{a}o (Brazil).

\bibliography{Ref.bib}
\end{document}